\documentclass[%
 aip,
 amsmath,amssymb,
 reprint,%
]{revtex4-1}

\usepackage{graphicx}
\usepackage{dcolumn}
\usepackage{bm}

\usepackage[utf8]{inputenc}
\usepackage[T1]{fontenc}
\usepackage{mathptmx}
\usepackage{etoolbox}

\makeatletter
\def\@email#1#2{%
 \endgroup
 \patchcmd{\titleblock@produce}
  {\frontmatter@RRAPformat}
  {\frontmatter@RRAPformat{\produce@RRAP{*#1\href{mailto:#2}{#2}}}\frontmatter@RRAPformat}
  {}{}
}%
\makeatother
\begin{document}

\preprint{AIP/123-QED}

\title{Femtosecond laser-shockwave induced densification in fused silica}
\author{A. Radhakrishnan}
 \email{arunkrishnan.radhakrishnan@epfl.ch}
 \affiliation{Galatea Laboratory, IMT/STI, Ecole Polytechnique Fédérale de Lausanne (EPFL), Rue de la Maladière 71b, 2000 Neuchâtel, Switzerland}
\author{J. Gateau}%
 \affiliation{Galatea Laboratory, IMT/STI, Ecole Polytechnique Fédérale de Lausanne (EPFL), Rue de la Maladière 71b, 2000 Neuchâtel, Switzerland} 
\author{P. Vlugter}
\affiliation{Galatea Laboratory, IMT/STI, Ecole Polytechnique Fédérale de Lausanne (EPFL), Rue de la Maladière 71b, 2000 Neuchâtel, Switzerland} 
\author{Y. Bellouard}
\affiliation{Galatea Laboratory, IMT/STI, Ecole Polytechnique Fédérale de Lausanne (EPFL), Rue de la Maladière 71b, 2000 Neuchâtel, Switzerland} 

\date{\today}
\begin{abstract}
Tightly focused femtosecond laser-beam in the non-ablative regime can induce a shock-wave enough to reach locally pressures in the giga-Pascal range or more. In a single beam configuration, the location of the highest-pressure zone is nested within the laser-focus zone, making it difficult to differentiate the effect of the shock-wave pressure from photo-induced and plasma relaxation effect. To circumvent this difficulty, we consider two spatially separated focused beams that individually act as quasi-simultaneous pressure-wave emitters. The zone where both shock-waves interfere constructively forms a region of extreme pressure range, physically separated from the regions under direct laser exposure. Here, we present evidences of pressured-induced densification in fused silica in between the foci of the two beams, which can be exclusively attributed to the superposition of the pressure waves emitted by each focused laser-beam. Specifically, we show how the beams gap and pulses time-delay affect the structural properties of fused silica using Raman characterization, beam deflection technique, and selective etching techniques. The method is generic and can be implemented in a variety of transparent substrates for high-pressure physics studies and, unlike classical methods, such as the use of diamond anvils, offers a means to create arbitrary-shaped laser-induced high-pressure impacted zones by scanning the two beams across the specimen volume.
\end{abstract}

\maketitle

\section{\label{sec:level1}Introduction}

\begin{figure}[ht]
    \centering
    \includegraphics[width=0.5\textwidth,height=\textheight,keepaspectratio]{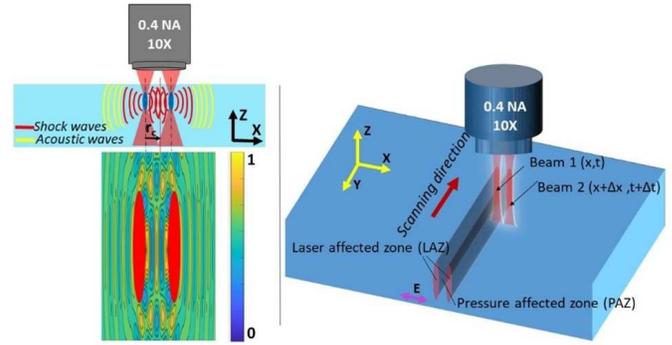}
    \caption{Left: Double-beam femtosecond laser bulk exposure: schematic view of the cross section (top left) and constructive wave interference pattern shown in the acoustic regime for illustrative purpose (bottom left). The two laser beams emit shock waves that add up in between the beams location. These decaying shock waves interfere constructively in between the two foci. Right: Continuous line-patterns scanning principle. The double-beam exposure can be applied anywhere through the specimens. The two pulses are temporally separated by the pulse length to prevent optical interferences, but yet for a duration order of magnitude shorter than the time scale of shock-wave emission to consider the two events as instantaneous.}
    \label{figure 1}
\end{figure}

In nature, high-pressure phases of silica are found in meteorite craters resulting from high-velocity impacts  \cite{gilletShockEventsSolar00}. Studying these phases in a laboratory setting remains a tedious task, as it requires high-pressure generation, from tens of giga-pascal (GPa) to tera-pascal (TPa) levels. While diamond anvil cells (DAC) are commonly used for the laboratory-scale high-pressure generation, it suffers from intrinsic limitations of volume and processing time restrictions \cite{jayaramanDiamondAnvilCell1983a,richetPressureinducedAmorphizationMinerals1997}. Due to the extreme brevity of the energy deposition and the rapid formation of a plasma, ultrafast laser interaction with dielectrics creates the conditions for locally achieving pressure levels in the TPa levels, as reported in sapphire and fused silica using single spot experiments  \cite{juodkazisLaserInducedMicroexplosionConfined2006,aEvidenceSuperdenseAluminium2011, gamalyAblationSolidsFemtosecond2002,juodkazisStructuralChangesFemtosecond2010}. There, the high-pressure zone is nested within the laser-affected zone, making it challenging to differentiate photo-induced from pressure-only effects as the material is not exclusively subjected to intense pressure waves, but also to the outcome related to plasma generated at the laser spot. 
To circumvent this difficulty and to effectively separate the high-pressure zone from the regions under direct exposure, we focus two spatially separated femtosecond. The two act as quasi-simultaneous emitters for strong pressure waves that interfere one another. In the case of constructive interferences, this configuration can lead to a higher-pressure zone located outside of the zone under direct laser exposure. When a femtosecond pulse is absorbed by the material, it forms shock waves caused by the rapid plasma volume formation, expansion, and decay  \cite{devauxGenerationShockWaves1993,zengLaserinducedShockwavePropagation2006,bertheShockWavesWaterconfined1997,hayasakiTimeresolvedInterferometryFemtosecondlaserinduced2011, bergnerSpatiotemporalAnalysisGlass2018,sakakuraObservationPressureWave2007}. These shock-waves propagate radially from the laser-propagation axis, at an initial velocity faster than the sound wave in the material (5.9x$10^3$ m/s for silica) and decays rapidly into acoustic sound waves within micron distances  \cite{gamalyLasermatterInteractionBulk2006}. Here, we increase the pressure by beam superposition technique as schematically illustrated in Figure \ref{figure 1}. (left). 

\section{\label{sec:level1}Experimental setup}

\begin{figure*}[ht]
    \centering
    \includegraphics[width=1\textwidth,height=\textheight,keepaspectratio]{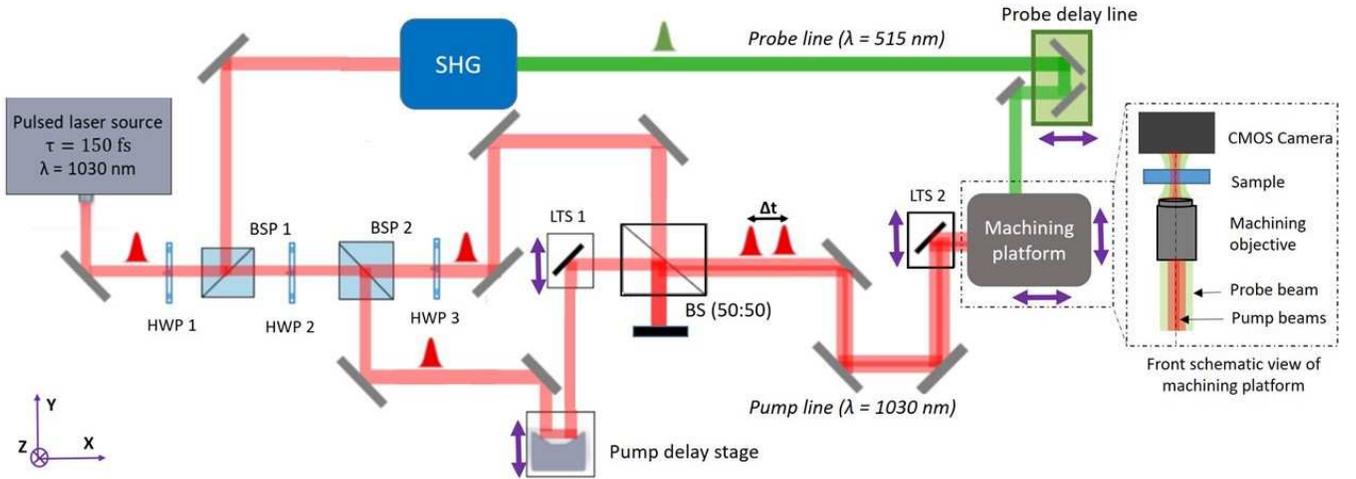}
    \caption{Optical layout of the double-beam experiment setup. A part of the incoming laser beam is converted into a frequency-doubled probe beam using a BBO crystal. The beam gap and the temporal delay are adjusted by a linear translation stage (LTS 1) and a pump delay line respectively. The front schematic sketch of the machining platform is shown in the right inset. (BSP: Beam splitter polarizer cube, HWP: Half wave plate, LTS: Linear translation stage, BBO: beta barium borate crystal.)}
    \label{figure 2}
\end{figure*}

Figure \ref{figure 2} illustrates the optical setup used in these experiments. A femtosecond laser (s-Pulse from Amplitude) delivering 150 fs pulses with a center wavelength of 1030 nm at a pulse repetition rate of 100 kHz is focused inside a fused silica substrate (Coring 7980 OF). The laser beam is split into two beams of equal energy that are temporally and spatially separated by a varying time delay ($\Delta$t) and beam gap ($\Delta$t), respectively. The two beams are focused with the same objective (OptoSigma, NA 0.4). A low-energy second harmonics probe pulse ($\lambda_{probe}$ = 515 nm), generated in a Beta Barium Borate (BBO) crystal is used to precisely control the time delay between the two incoming beams, with a resolution down to tens of femtoseconds between the pump and probe signals, by visually observing the effect of the plasma formation on a standard CMOS camera placed above the focusing objective. The spatial gap between the two machining beams is calibrated by optimizing the interference fringe pattern of the two beams and further refined by measuring the distance between laser-affected zones as observed in the SEM images of the polished cross-section of the written lines. In this set of experiments, we were able to vary the distance between laser-affected zones from 480 nm to 2 microns, and the time delay from zero to 66 ps. The machining was done in the bulk of the material at a distance of 20 microns from the surface.

\section{\label{sec:level1}results and discussions}
\subsection{\label{sec:level2}Densification measurements}

\begin{figure*}[ht]
    \centering
    \includegraphics[width=1\textwidth,height=\textheight,keepaspectratio]{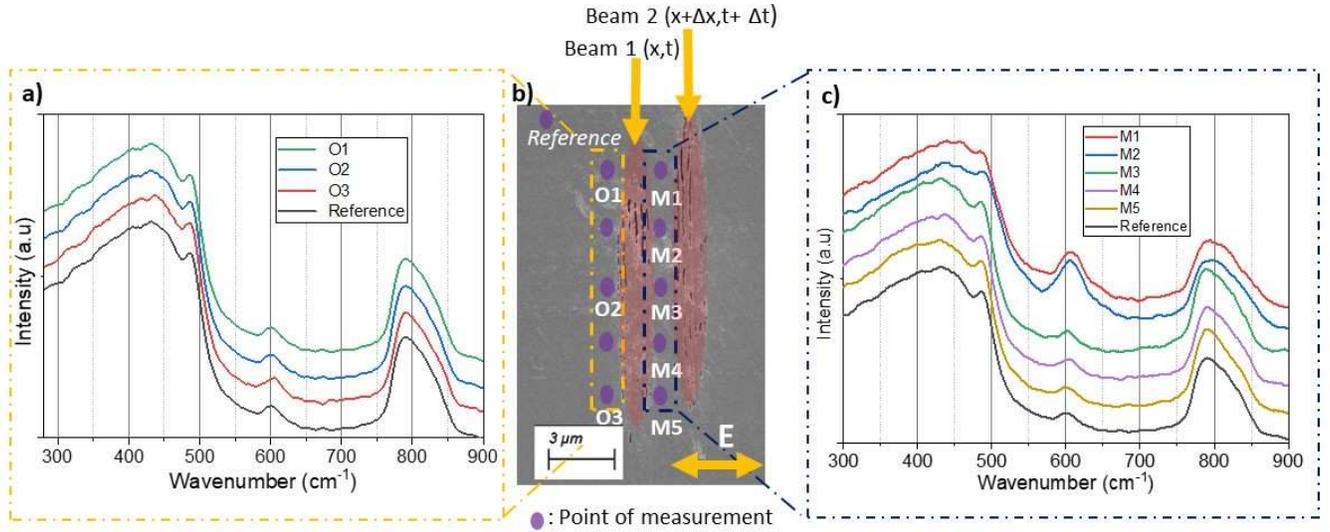}
    \caption{(middle) Microscopic image of the point of measurement along the cross-section of the machined lines; Left \& right / Raman spectra of the zones outside  the laser affected zone (denoted by O), and along the pressure affected zone in the middle of the beams (M) respectively.}
    \label{figure 3}
\end{figure*}

The cross section of the line-patterns produced by the double beams were analyzed using micro-Raman spectroscopy (Renishaw InVia confocal Raman spectroscope using a 100X objective, spot size < 1 $\mu$m). Here, we choose 405 nm as the excitation laser wavelength for stimulated Raman emission to prevent fluorescence excitation of laser-induced defects that could shadow other Raman peaks. Each measurement point in the scan was obtained with a laser power of 4.6 mW, and a total exposure time of 20 seconds. Figure \ref{figure 3}.b represents the Raman measurements taken outside and within the laser-modified zones, polished along their cross-section oriented towards the optical laser propagation axis. In this illustration, the beams are spatially separated by a distance of 1.2 microns and temporally, by a time delay of 240 fs. The data are presented for three characteristic zones, namely, the zone outside the laser affected zones (labelled ‘Zone O’), the zone located left to the laser exposed pattern (‘Zone L’), and finally, the zone in between the two laser affected zones (‘Zone M’), which is the zone where the two shockwaves are superposed. All Raman spectra are normalized with respect to the $\omega_4$ band, which is found to be more invariant to the effect of laser exposure. 
On one hand, ‘Zone O’, located one micron away from the laser exposed zone, shows no visible difference in the Raman spectra compared to a reference (as seen in Figure \ref{figure 3}.a) along all measurement points, while the pressure affected zone (zone M) was found to have modification in the region of shock superposition, in particularly among the points M1 and M2 as Figure \ref{figure 3}.c. Though we did not find a large variation in the shift of Raman spectra towards the higher wavenumbers (as reported in  \cite{okunoRamanSpectroscopicStudy1999}), we observe a rise of the D1 and D2 peak intensity (located around 495 cm$^{-1}$ and 605 cm$^{-1}$, respectively) \cite{galeenerBandLimitsVibrational1979,galeenerVibrationalDecouplingRings1984,galeenerLongitudinalOpticalVibrations1976}, along with a shift of the peaks towards higher wavenumbers as well as the shrinkage of the main band, which accounts for the reduction in bond angle of the silica lattice  \cite{senPhononsAXGlasses1977,sakakuraThermalShockInduced2011} in Zone M. 

\begin{figure*}[ht]
    \centering
    \includegraphics[width=1\textwidth,height=\textheight,keepaspectratio]{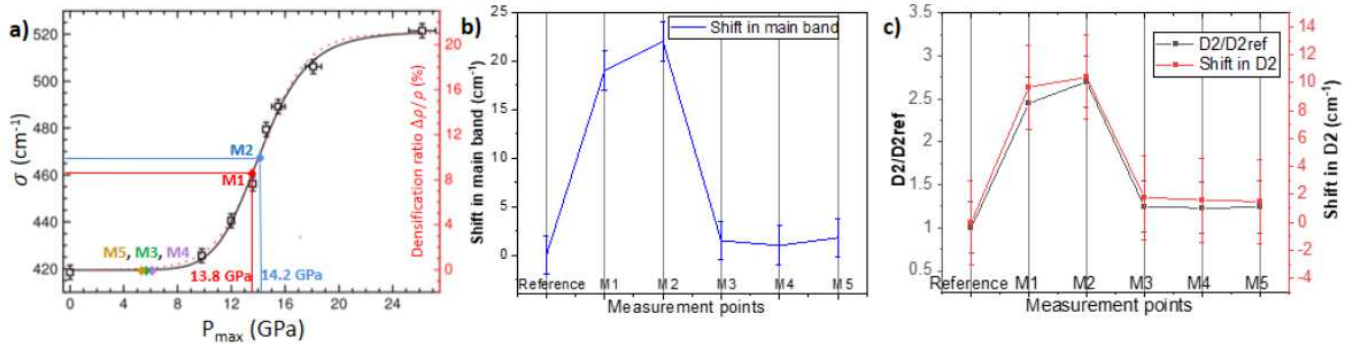}
    \caption{a. The Raman parameter $\sigma$ as a function of the maximum pressure reached $P_{max}$ as adapted from [\cite{sonnevilleProgressiveTransformationsSilica2012}] and the data from pressure affected zones was used to estimate the $P_{max}$ after double laser beam machining; b. The shift of the main band along the pressure affected zone in the middle of the beams (M); c. The shift of D2 peak and the ratio of the D2 peak with respect to the D2 reference peaks plotted along the pressure affected zone (M). }
    \label{figure 4}
\end{figure*}
These relative changes, combined in Figure \ref{figure 4}, point towards the densification of the material in between the two laser-affected zones. As this zone is located outside the laser-exposed zone, we attribute this densification to the effect of shock wave superposition. By comparing these variations in spectra to the compressive hydrostatic loading experiments reported elsewhere, we estimate a pressure development of about 13-14 GPa by estimating the Raman parameter $\sigma$ and the main band shift as mentioned in \cite{deschampsPermanentDensificationCompressed2013,sonnevilleProgressiveTransformationsSilica2012} (see Fig. \ref{figure 4} a). While this method gives a first estimate of the pressure, it assumes a hydrostatic loading case, which differs from our conditions and in fact, underestimates the real pressure. In our case, the modifications are obtained as a result of dynamic shock waves superposition. Based on silica shock-wave densification studies \cite{okunoRamanSpectroscopicStudy1999}, we estimate a pressure development of about 25-30 GPa by correlating the shift and rise of the D2 peak as shown in Figure \ref{figure 4}.c. The latter is of higher pressure since we have the shock superposition just over a limited period of time, and hence, higher pressure loading conditions. To explain why points M3-M5 do not show the same behavior as point M1-M2, we note that there was a shift in one of the beam along the optical propagation axis direction due to alignment errors, which may have resulted in uneven pressure distribution. Another possible explanation, as it will be further when examining Raman spectra \textit{inside }laser affected zones is the anisotropic pressure distribution in these zones.  These two observations may explain the lack of evidences for densified zones in between the tails of laser affected zones.

\subsubsection{\label{sec:level3}Modification along the laser affected zone}

\begin{figure*}[htp]
    \centering
    \includegraphics[width=1\textwidth,height=\textheight,keepaspectratio]{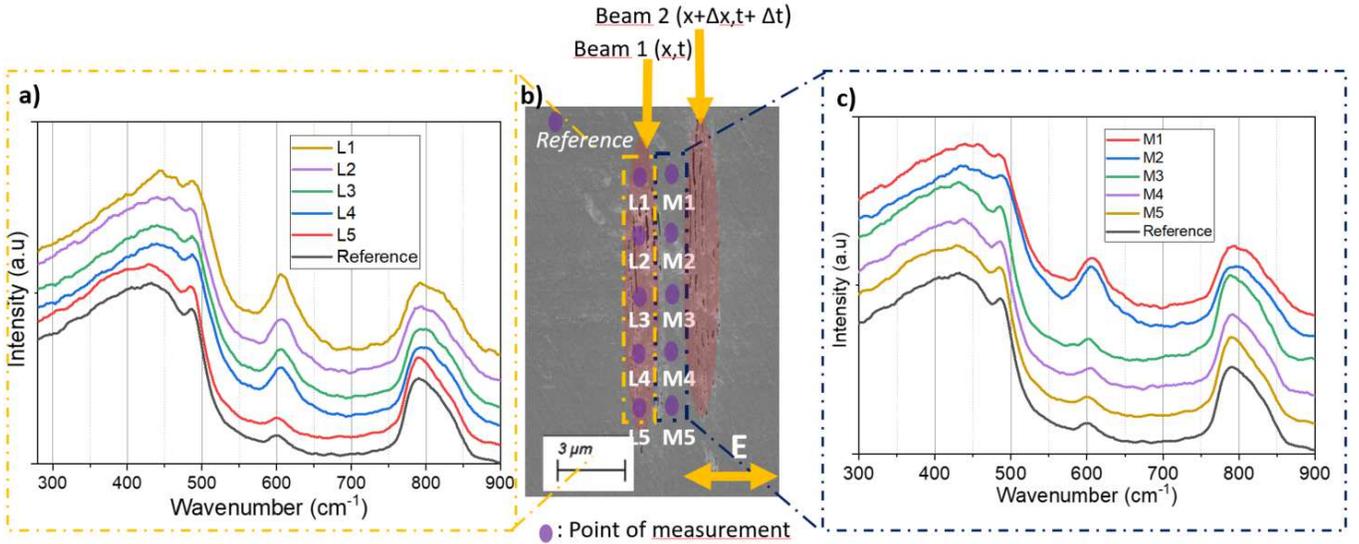}
    \caption{(middle) Microscopic image of the point of measurement along the cross-section of the machined lines; (Left and right) Raman spectra of the zones along the laser affected zone (denoted by L). The same Raman spectra shown for the region in between laser affected zones (denotes M) are shown for comparison. }
    \label{figure 5}
\end{figure*}
The ‘Zone L’, located in the left laser affected zone, exhibits clear modifications in the Raman spectra, which is expected based on previously reported data \cite{bellouardScanningThermalMicroscopy2008,chanStructuralChangesFused2001} as shown in Fig. \ref{figure 5}. (left), when compared to the pressure affected zone (Fig. \ref{figure 5}. right). Measurement points near the head of the laser affected zone show more densification in the Raman spectra than the ones measured near the tail of the LAZ. The tear-shaped geometry of the LAZ and the reduced intensity near the trailing edge \cite{docchioStudyTemporalSpatial1988,SpatialDistributionRefractive} may explain this effect. We observed a density of 2.35 g/cm$^3$, while comparing the D2 intensity and FWHM of the main band as mentioned in \cite{wangOverviewThermalErasure2021}. 

\begin{figure}[htp]
    \centering
    \includegraphics[width=0.5\textwidth,height=\textheight,keepaspectratio]{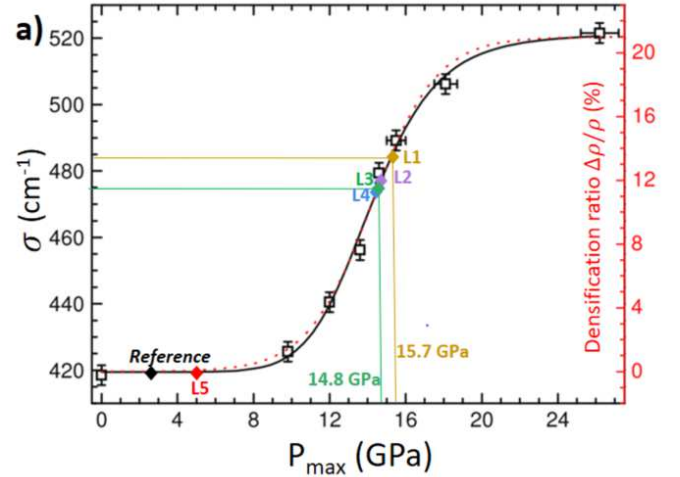}
    \caption{The Raman parameter $\sigma$ as a function of the maximum pressure reached $P_{max}$ as adapted from [\cite{deschampsPermanentDensificationCompressed2013}]. The data inside the laser affected zones (Zone L) are superimposed to estimate $P_{max}$ in our specific case.  }
    \label{figure 6}
\end{figure}

It was also found that the densification \textit{inside} the laser affected zone (LAZ) is higher than the densification due to homogeneous modification, a densification exposure regime found at shorter pulse duration and lower pulse energy \cite{Vlugter2022-ah}. Using the same method proposed by \cite{deschampsPermanentDensificationCompressed2013}  and used for a the first estimate of the pressure in between the beams, we estimated a pressure approaching ~15 GPa, see Fig \ref{figure 6}.This estimate is based on the assumption of an hydrostatic pressure and therefore, most likely underestimates what the real pressure was, as the loading case is present for a short duration and. here may liken the one of a shock wave.  However, as we are within the laser-affected zone, and hence in the zone where the plasma was located, it remains speculative to truly assess what the pressure conditions as we lack an equation of state for the matter under such extreme conditions. We would therefore caution that unlike the zone in between laser affected zones where there is a clear decoupling between plasma and pressurized zone, this pressure estimates based on Raman data performed \textit{inside} the laser-affected zones remain speculative. We also further noticed that the modification obtained inside the LAZ is independent from the presence of a shock wave emitted by a neighboring LAZ as identical Raman spectra were obtained inside LAZ, when the beams were far apart both spatially and temporally.

\subsection{\label{sec:level2}Effect of beam gap and time delay in the pressure affected zone}

\begin{figure*}[htp]
    \centering
    \includegraphics[width=1\textwidth,height=\textheight,keepaspectratio]{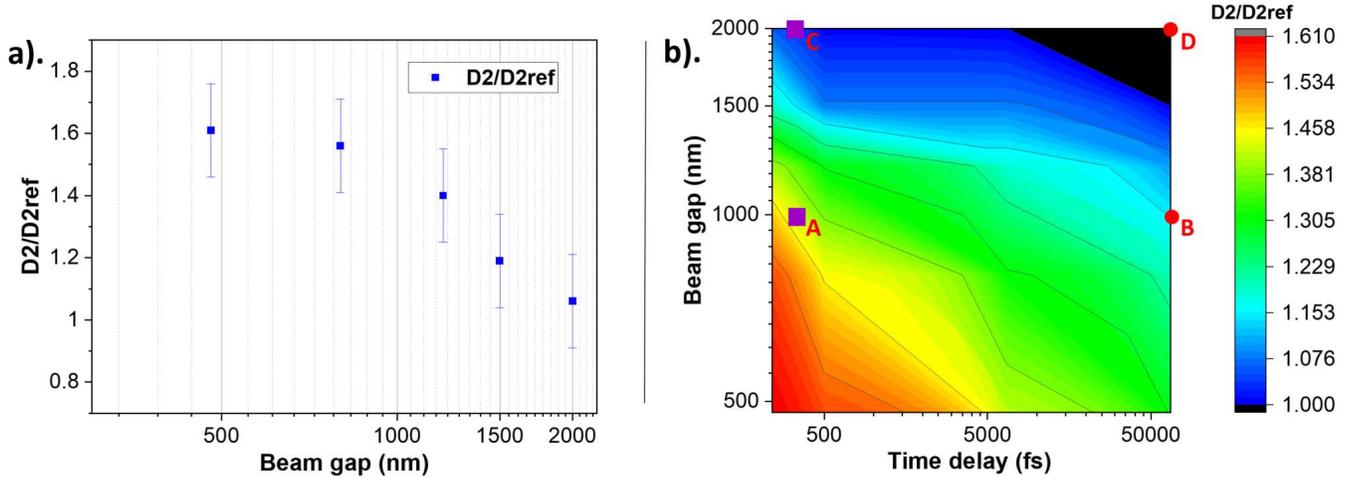}
    \caption{a. Comparison plot between D2/Dref ratio and beam gap ($\Delta$x) for a fixed time delay (240 fs); b. Contour map between beam gap and time delay in terms of the extent of the rise in D2 peak.}
    \label{figure 7}
\end{figure*}
To further investigate the densification effect, dual-line patterns were drawn in the fused silica substrates for varying beam gap ($\Delta$x) and time delay ($\Delta$t). The patterns were systematically analyzed using Raman spectroscopy following the same method than described in the previous paragraph and we used the ratio D2/Dref as an indirect metric of the densification. Figure \ref{figure 7} denotes the densification with varying beam gap, but for a constant time delay of 240 fs. The data corresponds to point M2 in Zone C, as indicated in Figure \ref{figure 7}.a, a point where the maximum densification was obtained. The maximum densification peaks between 480 nm and 1.2 microns, which indicates the expected decay of the pressure waves \cite{juodkazisLaserInducedMicroexplosionConfined2006,zeldovichPhysicsShockWaves2002} after a certain critical radius $r_c$ as shown in Figure \ref{figure 1}. (left). Beyond $r_c$, the pressure superposition is not sufficient to induce a permanent densification. Figure \ref{figure 7}.b. shows a contour plot for D2/ Dref contrast for varying time delays for various beam gaps. By taking into account the supersonic velocity approximation from Hugoniot data of fused silica  \cite{koenigOpticalSmoothingTechniques1994,desjarlaisExtensionHugoniotAnalytical2017}, we could suppose that above 6.6 ps, the superposition occurs, near or beyond the boundary of the trailing beam’s laser affected zone. Further, for lines exposed with ‘infinite time delay’, i.e. lines written sequentially one after the other, masking each beam sequentially we did not achieve a much higher densification parameters that the beams written without masking each other. The map suggests that maximum densification is obtained when the beam gap is less than one micron, and when the beam delay is less than 500 fs. This window of parameters can be used for generating a localized densified zone between two laser-affected zones, thereby having a varying refractive index regime. 

\subsection{\label{sec:level2} Volume variation measurement in double beam machined fused silica specimen as evidence of densification}

\begin{figure}[htp]
    \centering
    \includegraphics[width=0.48\textwidth,height=\textheight,keepaspectratio]{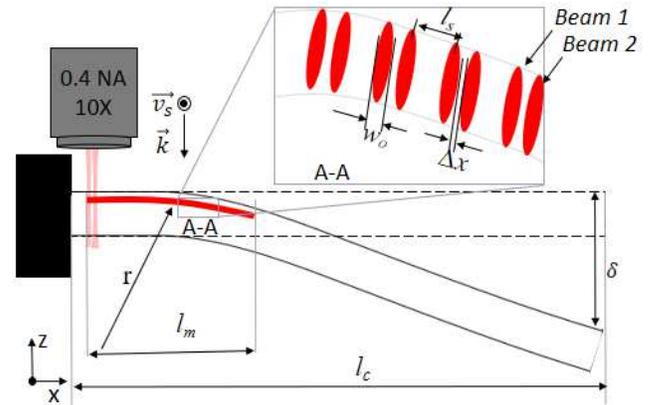}
    \caption{Schematic cross section of a cantilever, used with double-beam laser exposure conditions. In this illustration, the cantilever bends with a radius r, and results in an overall deflection  $\delta$, due to volume expansion. In the magnified rectangle A-A, the red ellipses represent the line cross-section, with individual width $w_0$ and separated by a beam gap $\Delta x$.}
    \label{figure 8}
\end{figure}
\begin{figure*}[htp]
    \centering
    \includegraphics[width=0.85\textwidth,height=\textheight,keepaspectratio]{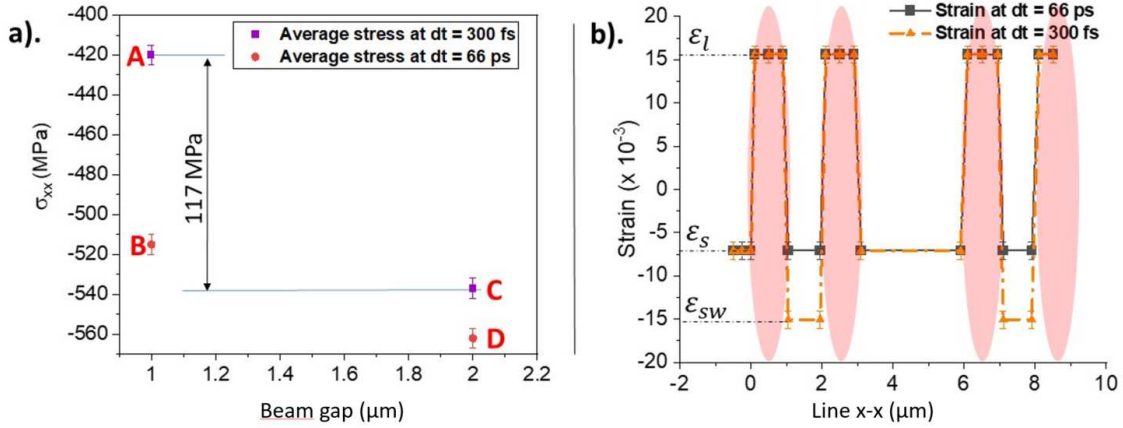}
    \caption{a. Average stress for selected time delays (dt) and beam gaps (A-D). Exposure conditions for these four cases are shown in Figure 4; b. Strain along an arc x-x for a beam gap of 1 micron, as shown in the inset of Figure 8.}
    \label{figure 9}
\end{figure*}

 To complement Raman data and to confirm the presence of densification, we use a method based on micro-cantilevers \cite{bellouardStressstateManipulationFused2016,vlugterElasticPropertiesSelforganized2020} to investigate localized volume variations resulting from the double-beam femtosecond laser exposure inside the bulk of silica. This method is highly sensitive and has also been used for investigating coefficient of thermal expansion changes after laser exposure. The working principle of this experimental technique, adapted to the double beam exposure, where a series of twin-lines (red zones in Fig. \ref{figure 8}) with a definite spacing ls are written along the upper part of the cantilever, and towards the anchoring joint. This results in a bimorph composite with an amplified displacement ($\delta$) in the transparent glass cantilever. Here, we exposed various cantilevers with varying time delays and beam gaps.The average stress and strain in the laser affected zones are extracted from the measured cantilever deflections using  Stoney's equation  \cite{championDirectVolumeVariation2012}. The results are shown in Figure \ref{figure 9} for four representative exposure conditions (labelled A  to D) and reported in Figure \ref{figure 7}.b. The difference in the average stress between case A, which is the one corresponding to the highest density case suggested by Raman observations, is in agreement with the formation of a high density amorphous (HDA) phase \cite{gleasonUltrafastVisualizationCrystallization2015,gleasonTimeresolvedDiffractionShockreleased2017} in between the laser affected zones. Indeed, this zone results in a reduced cantilever deflection, due to the volume compaction found in between laser affected zones, which in turn leads to a decrease in average stress.
 
Further, mathematical modeling was done assuming a uniform stress along the laser affected zones, to estimate the strain along an arc oriented towards axis x. The contributions of the different zones to the average strain are given in equation below, where $\epsilon_l$, $\epsilon_s$, and $\epsilon_{sw}$ are strain in the laser affected volume (Zone L), in between the two set of lines (Zone O) and the zone where the shockwaves superimpose (Zone M), respectively. The strain in zone O is due to the constant stress ($\sigma_{xx}$) in the LAZ and it is defined as $$\epsilon_s= \frac{\sigma_{xx}}{E.\epsilon_l}$$ is the strain in laser affected zone and is retrieved from the experiments where no densification due to shock wave is. $V_l$, $V_s$, and $V_{sw}$ are the respective volume fractions and they are defined as $V_l={2w_0}/{l_s}$, $V_s=1-{2w_0}/{l_s}$ and $V_{sw}={\Delta x}/{l_s}$. The average stress shall be written as,
$$\epsilon_{avg}=\epsilon_lV_l+\epsilon_sV_s+\epsilon_{sw}V_{sw}$$.
Here, we assume there is no shock-wave contribution when the laser beams are sufficiently temporarily separated i.e. $\epsilon_{sw}=0$.  Solving the average stress equation for the two extreme cases as shown in Figure 7.b, where there is limited contribution and maximum contribution from the shockwave, in the case of a time delay of 66 ps and 300 fs, respectively. Though we may note a strain of 10$\%$ from the Raman shift of D2 as mentioned in \cite{okunoRamanSpectroscopicStudy1999}; It should also be noted that the peak densification obtained at point M2 as in Figure \ref{figure 4} is diluted for these measurement as we estimate the average strain from the cantilever deflection. We shall note how the strain varies along the section and the change in the strain rate in between the laser-affected zone, clearly emphasizing the effect of a remnant strain due to the shockwave ($\epsilon_sw$), and hence the presence of a shock-induced densified zone. The strain obtained in our case is an average stress, and hence it is logical to have a lower value compared to the densification ratio obtained in shockwave experiments for which the entire element is subjected to uniform shock loading \cite{okunoRamanSpectroscopicStudy1999}. 

\subsection{\label{sec:level2} Effect of dual-beam exposure parameters on chemical etching selectivity}

\begin{figure*}[htp]
    \centering
    \includegraphics[width=1\textwidth,height=\textheight,keepaspectratio]{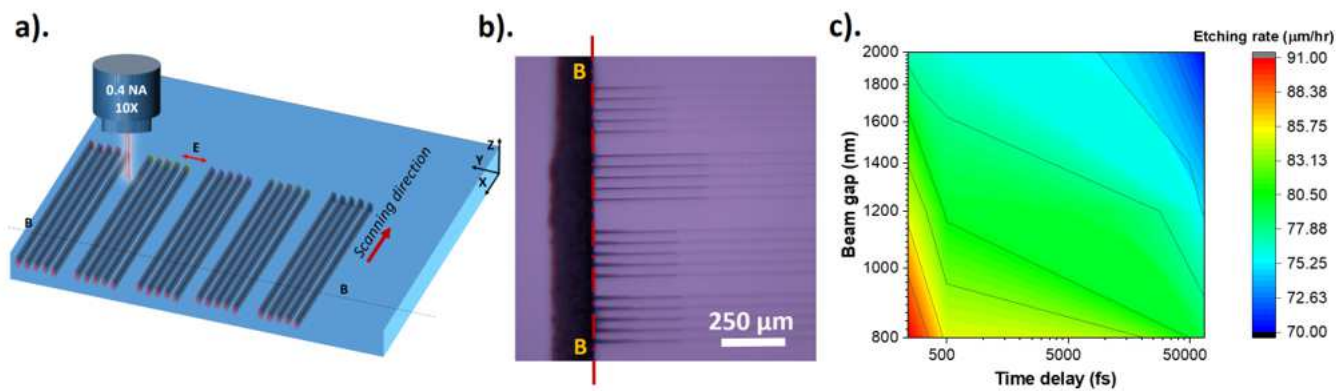}
    \caption{a. Schematic layout for the etching experiment. The sample is divided into several patterns, where each patch has a specific time delay and beam gap; b. The bright-field microscope top view image of selected patterns of the etched sample; c. Contour plot of etching rate with varying time delay and beam gap.}
    \label{figure 10}
\end{figure*}
It is known that femtosecond-laser affects the etching rate \cite{rajeshFastFemtosecondLaser2010,mouskeftarasEffectCombinationFemtosecond2018}, which itself depends on stress, density and structure of the materials. To understand the effect of dual-beam exposure on etching rate, a series of lines are drawn in the bulk of a fused silica substrate as shown Figure \ref{figure 10}.a. After exposure, the substrate is cut using a diamond wire-saw, along the section B-B as in the Figure \ref{figure 10}.b., and later submerged in 2.5\% HF bath for a preferential etching of 4.5 hours. The effect on etching rate of varying time delays and beam gaps is plotted as shown in Figure \ref{figure 10}.c.. The contour plot suggests a correlation with the densification experiments, and confirms other experiments reported in Agarwal \textit{et al}.  \cite{agarwalCorrelationSilicaGlass1997}. The increased etching rate is attributed to the densification due to the compressive loading by the shock waves that resulted in compressive stress similar to a mechanical, hydrostatic loading  \cite{agarwalCorrelationSilicaGlass1997}. 

\section{\label{sec:level1}Conclusion}

Using near simultaneous adjacent, but non-overlapping femtosecond laser beams, we have demonstrated that one can obtain permanent localized densification\textit{ in between} laser affected zones as the result of shock-wave superposition, and this, while preserving the integrity of the laser-exposed zones. Evidences of a localized densification were obtained using Raman spectroscopy, indirect remnant strain measurements and indirectly correlated with etching selectivity enhancement observations. Overall, the exposure method demonstrated here offers a means for studying the state of matter under high-velocity impact stress without the need for a direct contact method, like for instance the use of diamond anvil. Although, the pressure obtained remains moderate (a few tens of GPa), mainly due to the limit in laser power in our setup, this method demonstrates a path-way towards non-contact laser-induced high pressure studies where pressured-zone are separated from laser exposed zones. 

\subsection*{Data Availability Statement}
The data that support the findings of this study are available from the corresponding author upon reasonable request.
\begin{acknowledgments}
We acknowledge the Swiss National Science Foundation (FNS) for funding the Femto-Anvil project (Project number: FNS 200021 169681). We further acknowledge the financing form the ERC (ERC-2012-StG-307442). We thank Prof. Céline Hebert and Dr. Farhang Nabiei of the Physics Department, EPFL, and Ruben Ricca, Dr. Margarita Lesik Galatea lab, EPFL for the fruitful discussions and helping out with the cantilever experiment. We would also like to thank Dr. Richard Gaal and Dr. Arnoud Magrez of the Earth and planetary science department and Crystallographic facility, EPFL, respectively, for the training with the Raman spectroscope. Finally, the authors would like to thank Ruben Ricca for aiding in formatting the draft. 
\end{acknowledgments}

\nocite{*}
\bibliography{aipsamp}

\providecommand{\noopsort}[1]{}\providecommand{\singleletter}[1]{#1}
\begin{thebibliography}{39}%
\makeatletter
\providecommand \@ifxundefined [1]{%
 \@ifx{#1\undefined}
}%
\providecommand \@ifnum [1]{%
 \ifnum #1\expandafter \@firstoftwo
 \else \expandafter \@secondoftwo
 \fi
}%
\providecommand \@ifx [1]{%
 \ifx #1\expandafter \@firstoftwo
 \else \expandafter \@secondoftwo
 \fi
}%
\providecommand \natexlab [1]{#1}%
\providecommand \enquote  [1]{``#1''}%
\providecommand \bibnamefont  [1]{#1}%
\providecommand \bibfnamefont [1]{#1}%
\providecommand \citenamefont [1]{#1}%
\providecommand \href@noop [0]{\@secondoftwo}%
\providecommand \href [0]{\begingroup \@sanitize@url \@href}%
\providecommand \@href[1]{\@@startlink{#1}\@@href}%
\providecommand \@@href[1]{\endgroup#1\@@endlink}%
\providecommand \@sanitize@url [0]{\catcode `\\12\catcode `\$12\catcode
  `\&12\catcode `\#12\catcode `\^12\catcode `\_12\catcode `\%12\relax}%
\providecommand \@@startlink[1]{}%
\providecommand \@@endlink[0]{}%
\providecommand \url  [0]{\begingroup\@sanitize@url \@url }%
\providecommand \@url [1]{\endgroup\@href {#1}{\urlprefix }}%
\providecommand \urlprefix  [0]{URL }%
\providecommand \Eprint [0]{\href }%
\providecommand \doibase [0]{http://dx.doi.org/}%
\providecommand \selectlanguage [0]{\@gobble}%
\providecommand \bibinfo  [0]{\@secondoftwo}%
\providecommand \bibfield  [0]{\@secondoftwo}%
\providecommand \translation [1]{[#1]}%
\providecommand \BibitemOpen [0]{}%
\providecommand \bibitemStop [0]{}%
\providecommand \bibitemNoStop [0]{.\EOS\space}%
\providecommand \EOS [0]{\spacefactor3000\relax}%
\providecommand \BibitemShut  [1]{\csname bibitem#1\endcsname}%
\let\auto@bib@innerbib\@empty
\bibitem [{\citenamefont {Gillet}\ and\ \citenamefont
  {Goresy}(0)}]{gilletShockEventsSolar00}%
  \BibitemOpen
  \bibfield  {author} {\bibinfo {author} {\bibfnamefont {P.}~\bibnamefont
  {Gillet}}\ and\ \bibinfo {author} {\bibfnamefont {A.~E.}\ \bibnamefont
  {Goresy}},\ }\bibfield  {title} {\enquote {\bibinfo {title} {Shock {{Events}}
  in the {{Solar System}}: {{The Message}} from {{Minerals}} in {{Terrestrial
  Planets}} and {{Asteroids}}},}\ }\href {\doibase
  10.1146/annurev-ea-41-080913-200001} {\bibfield  {journal} {\bibinfo
  {journal} {Annual Review of Earth and Planetary Sciences}\ }\textbf {\bibinfo
  {volume} {0}},\ \bibinfo {pages} {257--285} (\bibinfo {year}
  {0})}\BibitemShut {NoStop}%
\bibitem [{\citenamefont {Jayaraman}(1983)}]{jayaramanDiamondAnvilCell1983a}%
  \BibitemOpen
  \bibfield  {author} {\bibinfo {author} {\bibfnamefont {A.}~\bibnamefont
  {Jayaraman}},\ }\bibfield  {title} {\enquote {\bibinfo {title} {Diamond anvil
  cell and high-pressure physical investigations},}\ }\href {\doibase
  10.1103/RevModPhys.55.65} {\bibfield  {journal} {\bibinfo  {journal} {Reviews
  of Modern Physics}\ }\textbf {\bibinfo {volume} {55}},\ \bibinfo {pages}
  {65--108} (\bibinfo {year} {1983})}\BibitemShut {NoStop}%
\bibitem [{\citenamefont {Richet}\ and\ \citenamefont
  {Gillet}(1997)}]{richetPressureinducedAmorphizationMinerals1997}%
  \BibitemOpen
  \bibfield  {author} {\bibinfo {author} {\bibfnamefont {P.}~\bibnamefont
  {Richet}}\ and\ \bibinfo {author} {\bibfnamefont {P.}~\bibnamefont
  {Gillet}},\ }\bibfield  {title} {\enquote {\bibinfo {title} {Pressure-induced
  amorphization of minerals; a review},}\ }\href {\doibase
  10.1127/EJM/9/5/0907} {\  (\bibinfo {year} {1997}),\
  10.1127/EJM/9/5/0907}\BibitemShut {NoStop}%
\bibitem [{\citenamefont {Juodkazis}\ \emph {et~al.}(2006)\citenamefont
  {Juodkazis}, \citenamefont {Nishimura}, \citenamefont {Tanaka}, \citenamefont
  {Misawa}, \citenamefont {Gamaly}, \citenamefont {{Luther-Davies}},
  \citenamefont {Hallo}, \citenamefont {Nicolai},\ and\ \citenamefont
  {Tikhonchuk}}]{juodkazisLaserInducedMicroexplosionConfined2006}%
  \BibitemOpen
  \bibfield  {author} {\bibinfo {author} {\bibfnamefont {S.}~\bibnamefont
  {Juodkazis}}, \bibinfo {author} {\bibfnamefont {K.}~\bibnamefont
  {Nishimura}}, \bibinfo {author} {\bibfnamefont {S.}~\bibnamefont {Tanaka}},
  \bibinfo {author} {\bibfnamefont {H.}~\bibnamefont {Misawa}}, \bibinfo
  {author} {\bibfnamefont {E.~G.}\ \bibnamefont {Gamaly}}, \bibinfo {author}
  {\bibfnamefont {B.}~\bibnamefont {{Luther-Davies}}}, \bibinfo {author}
  {\bibfnamefont {L.}~\bibnamefont {Hallo}}, \bibinfo {author} {\bibfnamefont
  {P.}~\bibnamefont {Nicolai}}, \ and\ \bibinfo {author} {\bibfnamefont
  {V.~T.}\ \bibnamefont {Tikhonchuk}},\ }\bibfield  {title} {\enquote {\bibinfo
  {title} {Laser-{{Induced Microexplosion Confined}} in the {{Bulk}} of a
  {{Sapphire Crystal}}: {{Evidence}} of {{Multimegabar Pressures}}},}\ }\href
  {\doibase 10.1103/PhysRevLett.96.166101} {\bibfield  {journal} {\bibinfo
  {journal} {Physical Review Letters}\ }\textbf {\bibinfo {volume} {96}},\
  \bibinfo {pages} {166101} (\bibinfo {year} {2006})}\BibitemShut {NoStop}%
\bibitem [{\citenamefont {A}\ \emph {et~al.}(2011)\citenamefont {A},
  \citenamefont {Eg}, \citenamefont {V}, \citenamefont {W}, \citenamefont
  {Av},\ and\ \citenamefont {S}}]{aEvidenceSuperdenseAluminium2011}%
  \BibitemOpen
  \bibfield  {author} {\bibinfo {author} {\bibfnamefont {V.}~\bibnamefont {A}},
  \bibinfo {author} {\bibfnamefont {G.}~\bibnamefont {Eg}}, \bibinfo {author}
  {\bibfnamefont {M.}~\bibnamefont {V}}, \bibinfo {author} {\bibfnamefont
  {Y.}~\bibnamefont {W}}, \bibinfo {author} {\bibfnamefont {R.}~\bibnamefont
  {Av}}, \ and\ \bibinfo {author} {\bibfnamefont {J.}~\bibnamefont {S}},\
  }\bibfield  {title} {\enquote {\bibinfo {title} {Evidence of superdense
  aluminium synthesized by ultrafast microexplosion.}}\ }\href {\doibase
  10.1038/ncomms1449} {\bibfield  {journal} {\bibinfo  {journal} {Nature
  Communications}\ }\textbf {\bibinfo {volume} {2}},\ \bibinfo {pages}
  {445--445} (\bibinfo {year} {2011})}\BibitemShut {NoStop}%
\bibitem [{\citenamefont {Gamaly}\ \emph {et~al.}(2002)\citenamefont {Gamaly},
  \citenamefont {Rode}, \citenamefont {{Luther-Davies}},\ and\ \citenamefont
  {Tikhonchuk}}]{gamalyAblationSolidsFemtosecond2002}%
  \BibitemOpen
  \bibfield  {author} {\bibinfo {author} {\bibfnamefont {E.~G.}\ \bibnamefont
  {Gamaly}}, \bibinfo {author} {\bibfnamefont {A.~V.}\ \bibnamefont {Rode}},
  \bibinfo {author} {\bibfnamefont {B.}~\bibnamefont {{Luther-Davies}}}, \ and\
  \bibinfo {author} {\bibfnamefont {V.~T.}\ \bibnamefont {Tikhonchuk}},\
  }\bibfield  {title} {\enquote {\bibinfo {title} {Ablation of solids by
  femtosecond lasers: {{Ablation}} mechanism and ablation thresholds for metals
  and dielectrics},}\ }\href {\doibase 10.1063/1.1447555} {\bibfield  {journal}
  {\bibinfo  {journal} {Physics of Plasmas}\ }\textbf {\bibinfo {volume} {9}},\
  \bibinfo {pages} {949--957} (\bibinfo {year} {2002})}\BibitemShut {NoStop}%
\bibitem [{\citenamefont {Juodkazis}\ \emph {et~al.}(2010)\citenamefont
  {Juodkazis}, \citenamefont {Kohara}, \citenamefont {Ohishi}, \citenamefont
  {Hirao}, \citenamefont {Vailionis}, \citenamefont {Mizeikis}, \citenamefont
  {Saito},\ and\ \citenamefont
  {Rode}}]{juodkazisStructuralChangesFemtosecond2010}%
  \BibitemOpen
  \bibfield  {author} {\bibinfo {author} {\bibfnamefont {S.}~\bibnamefont
  {Juodkazis}}, \bibinfo {author} {\bibfnamefont {S.}~\bibnamefont {Kohara}},
  \bibinfo {author} {\bibfnamefont {Y.}~\bibnamefont {Ohishi}}, \bibinfo
  {author} {\bibfnamefont {N.}~\bibnamefont {Hirao}}, \bibinfo {author}
  {\bibfnamefont {A.}~\bibnamefont {Vailionis}}, \bibinfo {author}
  {\bibfnamefont {V.}~\bibnamefont {Mizeikis}}, \bibinfo {author}
  {\bibfnamefont {A.}~\bibnamefont {Saito}}, \ and\ \bibinfo {author}
  {\bibfnamefont {A.}~\bibnamefont {Rode}},\ }\bibfield  {title} {\enquote
  {\bibinfo {title} {Structural changes in femtosecond laser modified regions
  inside fused silica},}\ }\href {\doibase 10.1088/2040-8978/12/12/124007}
  {\bibfield  {journal} {\bibinfo  {journal} {Journal of Optics}\ }\textbf
  {\bibinfo {volume} {12}},\ \bibinfo {pages} {124007} (\bibinfo {year}
  {2010})}\BibitemShut {NoStop}%
\bibitem [{\citenamefont {Devaux}\ \emph {et~al.}(1993)\citenamefont {Devaux},
  \citenamefont {Fabbro}, \citenamefont {Tollier},\ and\ \citenamefont
  {Bartnicki}}]{devauxGenerationShockWaves1993}%
  \BibitemOpen
  \bibfield  {author} {\bibinfo {author} {\bibfnamefont {D.}~\bibnamefont
  {Devaux}}, \bibinfo {author} {\bibfnamefont {R.}~\bibnamefont {Fabbro}},
  \bibinfo {author} {\bibfnamefont {L.}~\bibnamefont {Tollier}}, \ and\
  \bibinfo {author} {\bibfnamefont {E.}~\bibnamefont {Bartnicki}},\ }\bibfield
  {title} {\enquote {\bibinfo {title} {Generation of shock waves by
  laser-induced plasma in confined geometry},}\ }\href {\doibase
  10.1063/1.354710} {\bibfield  {journal} {\bibinfo  {journal} {Journal of
  Applied Physics}\ }\textbf {\bibinfo {volume} {74}},\ \bibinfo {pages}
  {2268--2273} (\bibinfo {year} {1993})}\BibitemShut {NoStop}%
\bibitem [{\citenamefont {Zeng}\ \emph {et~al.}(2006)\citenamefont {Zeng},
  \citenamefont {Mao}, \citenamefont {Mao}, \citenamefont {Wen}, \citenamefont
  {Greif},\ and\ \citenamefont
  {Russo}}]{zengLaserinducedShockwavePropagation2006}%
  \BibitemOpen
  \bibfield  {author} {\bibinfo {author} {\bibfnamefont {X.}~\bibnamefont
  {Zeng}}, \bibinfo {author} {\bibfnamefont {X.}~\bibnamefont {Mao}}, \bibinfo
  {author} {\bibfnamefont {S.~S.}\ \bibnamefont {Mao}}, \bibinfo {author}
  {\bibfnamefont {S.-B.}\ \bibnamefont {Wen}}, \bibinfo {author} {\bibfnamefont
  {R.}~\bibnamefont {Greif}}, \ and\ \bibinfo {author} {\bibfnamefont {R.~E.}\
  \bibnamefont {Russo}},\ }\bibfield  {title} {\enquote {\bibinfo {title}
  {Laser-induced shockwave propagation from ablation in a cavity},}\ }\href
  {\doibase 10.1063/1.2172738} {\bibfield  {journal} {\bibinfo  {journal}
  {Applied Physics Letters}\ }\textbf {\bibinfo {volume} {88}},\ \bibinfo
  {pages} {061502} (\bibinfo {year} {2006})}\BibitemShut {NoStop}%
\bibitem [{\citenamefont {Berthe}\ \emph {et~al.}(1997)\citenamefont {Berthe},
  \citenamefont {Fabbro}, \citenamefont {Peyre}, \citenamefont {Tollier},\ and\
  \citenamefont {Bartnicki}}]{bertheShockWavesWaterconfined1997}%
  \BibitemOpen
  \bibfield  {author} {\bibinfo {author} {\bibfnamefont {L.}~\bibnamefont
  {Berthe}}, \bibinfo {author} {\bibfnamefont {R.}~\bibnamefont {Fabbro}},
  \bibinfo {author} {\bibfnamefont {P.}~\bibnamefont {Peyre}}, \bibinfo
  {author} {\bibfnamefont {L.}~\bibnamefont {Tollier}}, \ and\ \bibinfo
  {author} {\bibfnamefont {E.}~\bibnamefont {Bartnicki}},\ }\bibfield  {title}
  {\enquote {\bibinfo {title} {Shock waves from a water-confined
  laser-generated plasma},}\ }\href {\doibase 10.1063/1.366113} {\bibfield
  {journal} {\bibinfo  {journal} {Journal of Applied Physics}\ }\textbf
  {\bibinfo {volume} {82}},\ \bibinfo {pages} {2826--2832} (\bibinfo {year}
  {1997})}\BibitemShut {NoStop}%
\bibitem [{\citenamefont {Hayasaki}\ \emph {et~al.}(2011)\citenamefont
  {Hayasaki}, \citenamefont {Isaka}, \citenamefont {Takita},\ and\
  \citenamefont
  {Juodkazis}}]{hayasakiTimeresolvedInterferometryFemtosecondlaserinduced2011}%
  \BibitemOpen
  \bibfield  {author} {\bibinfo {author} {\bibfnamefont {Y.}~\bibnamefont
  {Hayasaki}}, \bibinfo {author} {\bibfnamefont {M.}~\bibnamefont {Isaka}},
  \bibinfo {author} {\bibfnamefont {A.}~\bibnamefont {Takita}}, \ and\ \bibinfo
  {author} {\bibfnamefont {S.}~\bibnamefont {Juodkazis}},\ }\bibfield  {title}
  {\enquote {\bibinfo {title} {Time-resolved interferometry of
  femtosecond-laser-induced processes under tight focusing and close-to-optical
  breakdown inside borosilicate glass},}\ }\href {\doibase
  10.1364/OE.19.005725} {\bibfield  {journal} {\bibinfo  {journal} {Optics
  Express}\ }\textbf {\bibinfo {volume} {19}},\ \bibinfo {pages} {5725--5734}
  (\bibinfo {year} {2011})}\BibitemShut {NoStop}%
\bibitem [{\citenamefont {Bergner}\ \emph {et~al.}(2018)\citenamefont
  {Bergner}, \citenamefont {Seyfarth}, \citenamefont {Lammers}, \citenamefont
  {Ullsperger}, \citenamefont {D{\"o}ring}, \citenamefont {Heinrich},
  \citenamefont {Kumkar}, \citenamefont {Flamm}, \citenamefont
  {T{\"u}nnermann},\ and\ \citenamefont
  {Nolte}}]{bergnerSpatiotemporalAnalysisGlass2018}%
  \BibitemOpen
  \bibfield  {author} {\bibinfo {author} {\bibfnamefont {K.}~\bibnamefont
  {Bergner}}, \bibinfo {author} {\bibfnamefont {B.}~\bibnamefont {Seyfarth}},
  \bibinfo {author} {\bibfnamefont {K.~A.}\ \bibnamefont {Lammers}}, \bibinfo
  {author} {\bibfnamefont {T.}~\bibnamefont {Ullsperger}}, \bibinfo {author}
  {\bibfnamefont {S.}~\bibnamefont {D{\"o}ring}}, \bibinfo {author}
  {\bibfnamefont {M.}~\bibnamefont {Heinrich}}, \bibinfo {author}
  {\bibfnamefont {M.}~\bibnamefont {Kumkar}}, \bibinfo {author} {\bibfnamefont
  {D.}~\bibnamefont {Flamm}}, \bibinfo {author} {\bibfnamefont
  {A.}~\bibnamefont {T{\"u}nnermann}}, \ and\ \bibinfo {author} {\bibfnamefont
  {S.}~\bibnamefont {Nolte}},\ }\bibfield  {title} {\enquote {\bibinfo {title}
  {Spatio-temporal analysis of glass volume processing using ultrashort laser
  pulses},}\ }\href {\doibase 10.1364/AO.57.004618} {\bibfield  {journal}
  {\bibinfo  {journal} {Applied Optics}\ }\textbf {\bibinfo {volume} {57}},\
  \bibinfo {pages} {4618} (\bibinfo {year} {2018})}\BibitemShut {NoStop}%
\bibitem [{\citenamefont {Sakakura}\ \emph {et~al.}(2007)\citenamefont
  {Sakakura}, \citenamefont {Terazima}, \citenamefont {Shimotsuma},
  \citenamefont {Miura},\ and\ \citenamefont
  {Hirao}}]{sakakuraObservationPressureWave2007}%
  \BibitemOpen
  \bibfield  {author} {\bibinfo {author} {\bibfnamefont {M.}~\bibnamefont
  {Sakakura}}, \bibinfo {author} {\bibfnamefont {M.}~\bibnamefont {Terazima}},
  \bibinfo {author} {\bibfnamefont {Y.}~\bibnamefont {Shimotsuma}}, \bibinfo
  {author} {\bibfnamefont {K.}~\bibnamefont {Miura}}, \ and\ \bibinfo {author}
  {\bibfnamefont {K.}~\bibnamefont {Hirao}},\ }\bibfield  {title} {\enquote
  {\bibinfo {title} {Observation of pressure wave generated by focusing a
  femtosecond laser pulse inside a glass},}\ }\href {\doibase
  10.1364/OE.15.005674} {\bibfield  {journal} {\bibinfo  {journal} {Optics
  Express}\ }\textbf {\bibinfo {volume} {15}},\ \bibinfo {pages} {5674--5686}
  (\bibinfo {year} {2007})}\BibitemShut {NoStop}%
\bibitem [{\citenamefont {Gamaly}\ \emph {et~al.}(2006)\citenamefont {Gamaly},
  \citenamefont {Juodkazis}, \citenamefont {Nishimura}, \citenamefont {Misawa},
  \citenamefont {{Luther-Davies}}, \citenamefont {Hallo}, \citenamefont
  {Nicolai},\ and\ \citenamefont
  {Tikhonchuk}}]{gamalyLasermatterInteractionBulk2006}%
  \BibitemOpen
  \bibfield  {author} {\bibinfo {author} {\bibfnamefont {E.~G.}\ \bibnamefont
  {Gamaly}}, \bibinfo {author} {\bibfnamefont {S.}~\bibnamefont {Juodkazis}},
  \bibinfo {author} {\bibfnamefont {K.}~\bibnamefont {Nishimura}}, \bibinfo
  {author} {\bibfnamefont {H.}~\bibnamefont {Misawa}}, \bibinfo {author}
  {\bibfnamefont {B.}~\bibnamefont {{Luther-Davies}}}, \bibinfo {author}
  {\bibfnamefont {L.}~\bibnamefont {Hallo}}, \bibinfo {author} {\bibfnamefont
  {P.}~\bibnamefont {Nicolai}}, \ and\ \bibinfo {author} {\bibfnamefont
  {V.~T.}\ \bibnamefont {Tikhonchuk}},\ }\bibfield  {title} {\enquote {\bibinfo
  {title} {Laser-matter interaction in the bulk of a transparent solid:
  {{Confined}} microexplosion and void formation},}\ }\href {\doibase
  10.1103/PhysRevB.73.214101} {\bibfield  {journal} {\bibinfo  {journal}
  {Physical Review B}\ }\textbf {\bibinfo {volume} {73}},\ \bibinfo {pages}
  {214101} (\bibinfo {year} {2006})}\BibitemShut {NoStop}%
\bibitem [{\citenamefont {Okuno}\ \emph {et~al.}(1999)\citenamefont {Okuno},
  \citenamefont {Reynard}, \citenamefont {Shimada}, \citenamefont {Syono},\
  and\ \citenamefont {Willaime}}]{okunoRamanSpectroscopicStudy1999}%
  \BibitemOpen
  \bibfield  {author} {\bibinfo {author} {\bibfnamefont {M.}~\bibnamefont
  {Okuno}}, \bibinfo {author} {\bibfnamefont {B.}~\bibnamefont {Reynard}},
  \bibinfo {author} {\bibfnamefont {Y.}~\bibnamefont {Shimada}}, \bibinfo
  {author} {\bibfnamefont {Y.}~\bibnamefont {Syono}}, \ and\ \bibinfo {author}
  {\bibfnamefont {C.}~\bibnamefont {Willaime}},\ }\bibfield  {title} {\enquote
  {\bibinfo {title} {A {{Raman Spectroscopic Study}} of {{Shock-Wave
  Densification}} of {{Vitreous Silica}}},}\ }\href {\doibase
  10.1007/s002690050190} {\bibfield  {journal} {\bibinfo  {journal} {Physics
  and Chemistry of Minerals}\ }\textbf {\bibinfo {volume} {26}},\ \bibinfo
  {pages} {304--311} (\bibinfo {year} {1999})}\BibitemShut {NoStop}%
\bibitem [{\citenamefont {Galeener}(1979)}]{galeenerBandLimitsVibrational1979}%
  \BibitemOpen
  \bibfield  {author} {\bibinfo {author} {\bibfnamefont {F.~L.}\ \bibnamefont
  {Galeener}},\ }\bibfield  {title} {\enquote {\bibinfo {title} {Band limits
  and the vibrational spectra of tetrahedral glasses},}\ }\href {\doibase
  10.1103/PhysRevB.19.4292} {\bibfield  {journal} {\bibinfo  {journal}
  {Physical Review B}\ }\textbf {\bibinfo {volume} {19}},\ \bibinfo {pages}
  {4292--4297} (\bibinfo {year} {1979})}\BibitemShut {NoStop}%
\bibitem [{\citenamefont {Galeener}\ \emph {et~al.}(1984)\citenamefont
  {Galeener}, \citenamefont {Barrio}, \citenamefont {Martinez},\ and\
  \citenamefont {Elliott}}]{galeenerVibrationalDecouplingRings1984}%
  \BibitemOpen
  \bibfield  {author} {\bibinfo {author} {\bibfnamefont {F.~L.}\ \bibnamefont
  {Galeener}}, \bibinfo {author} {\bibfnamefont {R.~A.}\ \bibnamefont
  {Barrio}}, \bibinfo {author} {\bibfnamefont {E.}~\bibnamefont {Martinez}}, \
  and\ \bibinfo {author} {\bibfnamefont {R.~J.}\ \bibnamefont {Elliott}},\
  }\bibfield  {title} {\enquote {\bibinfo {title} {Vibrational {{Decoupling}}
  of {{Rings}} in {{Amorphous Solids}}},}\ }\href {\doibase
  10.1103/PhysRevLett.53.2429} {\bibfield  {journal} {\bibinfo  {journal}
  {Physical Review Letters}\ }\textbf {\bibinfo {volume} {53}},\ \bibinfo
  {pages} {2429--2432} (\bibinfo {year} {1984})}\BibitemShut {NoStop}%
\bibitem [{\citenamefont {Galeener}\ and\ \citenamefont
  {Lucovsky}(1976)}]{galeenerLongitudinalOpticalVibrations1976}%
  \BibitemOpen
  \bibfield  {author} {\bibinfo {author} {\bibfnamefont {F.~L.}\ \bibnamefont
  {Galeener}}\ and\ \bibinfo {author} {\bibfnamefont {G.}~\bibnamefont
  {Lucovsky}},\ }\bibfield  {title} {\enquote {\bibinfo {title} {Longitudinal
  optical vibrations in glasses: Geo2 and sio2},}\ }\href {\doibase
  10.1103/PhysRevLett.37.1474} {\bibfield  {journal} {\bibinfo  {journal}
  {Physical Review Letters}\ }\textbf {\bibinfo {volume} {37}},\ \bibinfo
  {pages} {1474--1478} (\bibinfo {year} {1976})}\BibitemShut {NoStop}%
\bibitem [{\citenamefont {Sen}\ and\ \citenamefont
  {Thorpe}(1977)}]{senPhononsAXGlasses1977}%
  \BibitemOpen
  \bibfield  {author} {\bibinfo {author} {\bibfnamefont {P.~N.}\ \bibnamefont
  {Sen}}\ and\ \bibinfo {author} {\bibfnamefont {M.~F.}\ \bibnamefont
  {Thorpe}},\ }\bibfield  {title} {\enquote {\bibinfo {title} {Phonons in
  \${{A}}\{\vphantom\}{{X}}\vphantom\{\}\_\{2\}\$ glasses: {{From}} molecular
  to band-like modes},}\ }\href {\doibase 10.1103/PhysRevB.15.4030} {\bibfield
  {journal} {\bibinfo  {journal} {Physical Review B}\ }\textbf {\bibinfo
  {volume} {15}},\ \bibinfo {pages} {4030--4038} (\bibinfo {year}
  {1977})}\BibitemShut {NoStop}%
\bibitem [{\citenamefont {Sakakura}\ \emph {et~al.}(2011)\citenamefont
  {Sakakura}, \citenamefont {Terazima}, \citenamefont {Shimotsuma},
  \citenamefont {Miura},\ and\ \citenamefont
  {Hirao}}]{sakakuraThermalShockInduced2011}%
  \BibitemOpen
  \bibfield  {author} {\bibinfo {author} {\bibfnamefont {M.}~\bibnamefont
  {Sakakura}}, \bibinfo {author} {\bibfnamefont {M.}~\bibnamefont {Terazima}},
  \bibinfo {author} {\bibfnamefont {Y.}~\bibnamefont {Shimotsuma}}, \bibinfo
  {author} {\bibfnamefont {K.}~\bibnamefont {Miura}}, \ and\ \bibinfo {author}
  {\bibfnamefont {K.}~\bibnamefont {Hirao}},\ }\bibfield  {title} {\enquote
  {\bibinfo {title} {Thermal and shock induced modification inside a silica
  glass by focused femtosecond laser pulse},}\ }\href {\doibase
  10.1063/1.3533431} {\bibfield  {journal} {\bibinfo  {journal} {Journal of
  Applied Physics}\ }\textbf {\bibinfo {volume} {109}},\ \bibinfo {pages}
  {023503} (\bibinfo {year} {2011})}\BibitemShut {NoStop}%
\bibitem [{\citenamefont {Sonneville}\ \emph {et~al.}(2012)\citenamefont
  {Sonneville}, \citenamefont {Mermet}, \citenamefont {Champagnon},
  \citenamefont {Martinet}, \citenamefont {Margueritat}, \citenamefont {{de
  Ligny}}, \citenamefont {Deschamps},\ and\ \citenamefont
  {Balima}}]{sonnevilleProgressiveTransformationsSilica2012}%
  \BibitemOpen
  \bibfield  {author} {\bibinfo {author} {\bibfnamefont {C.}~\bibnamefont
  {Sonneville}}, \bibinfo {author} {\bibfnamefont {A.}~\bibnamefont {Mermet}},
  \bibinfo {author} {\bibfnamefont {B.}~\bibnamefont {Champagnon}}, \bibinfo
  {author} {\bibfnamefont {C.}~\bibnamefont {Martinet}}, \bibinfo {author}
  {\bibfnamefont {J.}~\bibnamefont {Margueritat}}, \bibinfo {author}
  {\bibfnamefont {D.}~\bibnamefont {{de Ligny}}}, \bibinfo {author}
  {\bibfnamefont {T.}~\bibnamefont {Deschamps}}, \ and\ \bibinfo {author}
  {\bibfnamefont {F.}~\bibnamefont {Balima}},\ }\bibfield  {title} {\enquote
  {\bibinfo {title} {Progressive transformations of silica glass upon
  densification},}\ }\href {\doibase 10.1063/1.4754601} {\bibfield  {journal}
  {\bibinfo  {journal} {The Journal of Chemical Physics}\ }\textbf {\bibinfo
  {volume} {137}},\ \bibinfo {pages} {124505} (\bibinfo {year}
  {2012})}\BibitemShut {NoStop}%
\bibitem [{\citenamefont {Deschamps}\ \emph {et~al.}(2013)\citenamefont
  {Deschamps}, \citenamefont {{Kassir-Bodon}}, \citenamefont {Sonneville},
  \citenamefont {Margueritat}, \citenamefont {Martinet}, \citenamefont {{de
  Ligny}}, \citenamefont {Mermet},\ and\ \citenamefont
  {Champagnon}}]{deschampsPermanentDensificationCompressed2013}%
  \BibitemOpen
  \bibfield  {author} {\bibinfo {author} {\bibfnamefont {T.}~\bibnamefont
  {Deschamps}}, \bibinfo {author} {\bibfnamefont {A.}~\bibnamefont
  {{Kassir-Bodon}}}, \bibinfo {author} {\bibfnamefont {C.}~\bibnamefont
  {Sonneville}}, \bibinfo {author} {\bibfnamefont {J.}~\bibnamefont
  {Margueritat}}, \bibinfo {author} {\bibfnamefont {C.}~\bibnamefont
  {Martinet}}, \bibinfo {author} {\bibfnamefont {D.}~\bibnamefont {{de
  Ligny}}}, \bibinfo {author} {\bibfnamefont {A.}~\bibnamefont {Mermet}}, \
  and\ \bibinfo {author} {\bibfnamefont {B.}~\bibnamefont {Champagnon}},\
  }\bibfield  {title} {\enquote {\bibinfo {title} {Permanent densification of
  compressed silica glass: A {{Raman-density}} calibration curve},}\ }\href
  {\doibase 10.1088/0953-8984/25/2/025402} {\bibfield  {journal} {\bibinfo
  {journal} {Journal of Physics: Condensed Matter}\ }\textbf {\bibinfo {volume}
  {25}},\ \bibinfo {pages} {025402} (\bibinfo {year} {2013})}\BibitemShut
  {NoStop}%
\bibitem [{\citenamefont {Bellouard}\ \emph {et~al.}(2008)\citenamefont
  {Bellouard}, \citenamefont {Barthel}, \citenamefont {Said}, \citenamefont
  {Dugan},\ and\ \citenamefont
  {Bado}}]{bellouardScanningThermalMicroscopy2008}%
  \BibitemOpen
  \bibfield  {author} {\bibinfo {author} {\bibfnamefont {Y.}~\bibnamefont
  {Bellouard}}, \bibinfo {author} {\bibfnamefont {E.}~\bibnamefont {Barthel}},
  \bibinfo {author} {\bibfnamefont {A.~A.}\ \bibnamefont {Said}}, \bibinfo
  {author} {\bibfnamefont {M.}~\bibnamefont {Dugan}}, \ and\ \bibinfo {author}
  {\bibfnamefont {P.}~\bibnamefont {Bado}},\ }\bibfield  {title} {\enquote
  {\bibinfo {title} {Scanning thermal microscopy and {{Raman}} analysis of bulk
  fused silica exposed to low-energy femtosecond laser pulses},}\ }\href
  {\doibase 10.1364/OE.16.019520} {\bibfield  {journal} {\bibinfo  {journal}
  {Optics Express}\ }\textbf {\bibinfo {volume} {16}},\ \bibinfo {pages}
  {19520--19534} (\bibinfo {year} {2008})}\BibitemShut {NoStop}%
\bibitem [{\citenamefont {Chan}\ \emph {et~al.}(2001)\citenamefont {Chan},
  \citenamefont {Huser}, \citenamefont {Risbud},\ and\ \citenamefont
  {Krol}}]{chanStructuralChangesFused2001}%
  \BibitemOpen
  \bibfield  {author} {\bibinfo {author} {\bibfnamefont {J.~W.}\ \bibnamefont
  {Chan}}, \bibinfo {author} {\bibfnamefont {T.}~\bibnamefont {Huser}},
  \bibinfo {author} {\bibfnamefont {S.}~\bibnamefont {Risbud}}, \ and\ \bibinfo
  {author} {\bibfnamefont {D.~M.}\ \bibnamefont {Krol}},\ }\bibfield  {title}
  {\enquote {\bibinfo {title} {Structural changes in fused silica after
  exposure to focused femtosecond laser pulses},}\ }\href {\doibase
  10.1364/OL.26.001726} {\bibfield  {journal} {\bibinfo  {journal} {Optics
  Letters}\ }\textbf {\bibinfo {volume} {26}},\ \bibinfo {pages} {1726--1728}
  (\bibinfo {year} {2001})}\BibitemShut {NoStop}%
\bibitem [{\citenamefont {Docchio}\ \emph {et~al.}(1988)\citenamefont
  {Docchio}, \citenamefont {Regondi}, \citenamefont {Capon},\ and\
  \citenamefont {Mellerio}}]{docchioStudyTemporalSpatial1988}%
  \BibitemOpen
  \bibfield  {author} {\bibinfo {author} {\bibfnamefont {F.}~\bibnamefont
  {Docchio}}, \bibinfo {author} {\bibfnamefont {P.}~\bibnamefont {Regondi}},
  \bibinfo {author} {\bibfnamefont {M.~R.~C.}\ \bibnamefont {Capon}}, \ and\
  \bibinfo {author} {\bibfnamefont {J.}~\bibnamefont {Mellerio}},\ }\bibfield
  {title} {\enquote {\bibinfo {title} {Study of the temporal and spatial
  dynamics of plasmas induced in liquids by nanosecond {{Nd}}:{{YAG}} laser
  pulses. 1: {{Analysis}} of the plasma starting times},}\ }\href {\doibase
  10.1364/AO.27.003661} {\bibfield  {journal} {\bibinfo  {journal} {Applied
  Optics}\ }\textbf {\bibinfo {volume} {27}},\ \bibinfo {pages} {3661--3668}
  (\bibinfo {year} {1988})}\BibitemShut {NoStop}%
\bibitem [{Spa()}]{SpatialDistributionRefractive}%
  \BibitemOpen
  \href@noop {} {\enquote {\bibinfo {title} {Spatial distribution of refractive
  index variations induced in bulk fused silica by single ultrashort and short
  laser pulses: {{Journal}} of {{Applied Physics}}: {{Vol}} 101, {{No}} 4},}\
  }\bibinfo {howpublished}
  {https://aip.scitation.org/doi/full/10.1063/1.2436925}\BibitemShut {NoStop}%
\bibitem [{\citenamefont {Wang}\ \emph {et~al.}(2021)\citenamefont {Wang},
  \citenamefont {Cavillon}, \citenamefont {Ollier}, \citenamefont {Poumellec},\
  and\ \citenamefont {Lancry}}]{wangOverviewThermalErasure2021}%
  \BibitemOpen
  \bibfield  {author} {\bibinfo {author} {\bibfnamefont {Y.}~\bibnamefont
  {Wang}}, \bibinfo {author} {\bibfnamefont {M.}~\bibnamefont {Cavillon}},
  \bibinfo {author} {\bibfnamefont {N.}~\bibnamefont {Ollier}}, \bibinfo
  {author} {\bibfnamefont {B.}~\bibnamefont {Poumellec}}, \ and\ \bibinfo
  {author} {\bibfnamefont {M.}~\bibnamefont {Lancry}},\ }\bibfield  {title}
  {\enquote {\bibinfo {title} {An {{Overview}} of the {{Thermal Erasure
  Mechanisms}} of {{Femtosecond Laser-Induced Nanogratings}} in {{Silica
  Glass}}},}\ }\href {\doibase 10.1002/pssa.202100023} {\bibfield  {journal}
  {\bibinfo  {journal} {physica status solidi (a)}\ }\textbf {\bibinfo {volume}
  {218}},\ \bibinfo {pages} {2100023} (\bibinfo {year} {2021})}\BibitemShut
  {NoStop}%
\bibitem [{\citenamefont {Vlugter}\ and\ \citenamefont
  {Bellouard}(2022)}]{Vlugter2022-ah}%
  \BibitemOpen
  \bibfield  {author} {\bibinfo {author} {\bibfnamefont {P.}~\bibnamefont
  {Vlugter}}\ and\ \bibinfo {author} {\bibfnamefont {Y.}~\bibnamefont
  {Bellouard}},\ }\bibfield  {title} {\enquote {\bibinfo {title} {On the
  abnormal temperature dependent elastic properties of fused silica irradiated
  by ultrafast lasers},}\ }\href@noop {} {\  (\bibinfo {year} {2022})},\
  \Eprint {http://arxiv.org/abs/2202.01886} {arXiv:2202.01886
  [cond-mat.mtrl-sci]} \BibitemShut {NoStop}%
\bibitem [{\citenamefont {Zel'dovich}\ and\ \citenamefont
  {Raizer}(2002)}]{zeldovichPhysicsShockWaves2002}%
  \BibitemOpen
  \bibfield  {author} {\bibinfo {author} {\bibfnamefont {Y.~B.}\ \bibnamefont
  {Zel'dovich}}\ and\ \bibinfo {author} {\bibfnamefont {Y.~P.}\ \bibnamefont
  {Raizer}},\ }\href@noop {} {\emph {\bibinfo {title} {Physics of {{Shock
  Waves}} and {{High-Temperature Hydrodynamic Phenomena}}}}},\ \bibinfo
  {edition} {annotated edition}\ ed.\ (\bibinfo  {publisher} {{Dover
  Publications}},\ \bibinfo {address} {{Mineola, N.Y}},\ \bibinfo {year}
  {2002})\BibitemShut {NoStop}%
\bibitem [{\citenamefont {Koenig}\ \emph {et~al.}(1994)\citenamefont {Koenig},
  \citenamefont {Faral}, \citenamefont {Boudenne}, \citenamefont {Batani},
  \citenamefont {Benuzzi},\ and\ \citenamefont
  {Bossi}}]{koenigOpticalSmoothingTechniques1994}%
  \BibitemOpen
  \bibfield  {author} {\bibinfo {author} {\bibfnamefont {M.}~\bibnamefont
  {Koenig}}, \bibinfo {author} {\bibfnamefont {B.}~\bibnamefont {Faral}},
  \bibinfo {author} {\bibfnamefont {J.~M.}\ \bibnamefont {Boudenne}}, \bibinfo
  {author} {\bibfnamefont {D.}~\bibnamefont {Batani}}, \bibinfo {author}
  {\bibfnamefont {A.}~\bibnamefont {Benuzzi}}, \ and\ \bibinfo {author}
  {\bibfnamefont {S.}~\bibnamefont {Bossi}},\ }\bibfield  {title} {\enquote
  {\bibinfo {title} {Optical smoothing techniques for shock wave generation in
  laser-produced plasmas},}\ }\href {\doibase 10.1103/PhysRevE.50.R3314}
  {\bibfield  {journal} {\bibinfo  {journal} {Physical Review E}\ }\textbf
  {\bibinfo {volume} {50}},\ \bibinfo {pages} {R3314--R3317} (\bibinfo {year}
  {1994})}\BibitemShut {NoStop}%
\bibitem [{\citenamefont {Desjarlais}, \citenamefont {Knudson},\ and\
  \citenamefont {Cochrane}(2017)}]{desjarlaisExtensionHugoniotAnalytical2017}%
  \BibitemOpen
  \bibfield  {author} {\bibinfo {author} {\bibfnamefont {M.~P.}\ \bibnamefont
  {Desjarlais}}, \bibinfo {author} {\bibfnamefont {M.~D.}\ \bibnamefont
  {Knudson}}, \ and\ \bibinfo {author} {\bibfnamefont {K.~R.}\ \bibnamefont
  {Cochrane}},\ }\bibfield  {title} {\enquote {\bibinfo {title} {Extension of
  the {{Hugoniot}} and analytical release model of {$\alpha$}-quartz to
  0.2\textendash 3 {{TPa}}},}\ }\href {\doibase 10.1063/1.4991814} {\bibfield
  {journal} {\bibinfo  {journal} {Journal of Applied Physics}\ }\textbf
  {\bibinfo {volume} {122}},\ \bibinfo {pages} {035903} (\bibinfo {year}
  {2017})}\BibitemShut {NoStop}%
\bibitem [{\citenamefont {Bellouard}\ \emph {et~al.}(2016)\citenamefont
  {Bellouard}, \citenamefont {Champion}, \citenamefont {McMillen},
  \citenamefont {Mukherjee}, \citenamefont {Thomson}, \citenamefont
  {P{\'e}pin}, \citenamefont {Gillet},\ and\ \citenamefont
  {Cheng}}]{bellouardStressstateManipulationFused2016}%
  \BibitemOpen
  \bibfield  {author} {\bibinfo {author} {\bibfnamefont {Y.}~\bibnamefont
  {Bellouard}}, \bibinfo {author} {\bibfnamefont {A.}~\bibnamefont {Champion}},
  \bibinfo {author} {\bibfnamefont {B.}~\bibnamefont {McMillen}}, \bibinfo
  {author} {\bibfnamefont {S.}~\bibnamefont {Mukherjee}}, \bibinfo {author}
  {\bibfnamefont {R.~R.}\ \bibnamefont {Thomson}}, \bibinfo {author}
  {\bibfnamefont {C.}~\bibnamefont {P{\'e}pin}}, \bibinfo {author}
  {\bibfnamefont {P.}~\bibnamefont {Gillet}}, \ and\ \bibinfo {author}
  {\bibfnamefont {Y.}~\bibnamefont {Cheng}},\ }\bibfield  {title} {\enquote
  {\bibinfo {title} {Stress-state manipulation in fused silica via femtosecond
  laser irradiation},}\ }\href {\doibase 10.1364/OPTICA.3.001285} {\bibfield
  {journal} {\bibinfo  {journal} {Optica}\ }\textbf {\bibinfo {volume} {3}},\
  \bibinfo {pages} {1285--1293} (\bibinfo {year} {2016})}\BibitemShut {NoStop}%
\bibitem [{\citenamefont {VLUGTER}\ and\ \citenamefont
  {BELLOUARD}(2020)}]{vlugterElasticPropertiesSelforganized2020}%
  \BibitemOpen
  \bibfield  {author} {\bibinfo {author} {\bibfnamefont {P.}~\bibnamefont
  {VLUGTER}}\ and\ \bibinfo {author} {\bibfnamefont {Y.}~\bibnamefont
  {BELLOUARD}},\ }\bibfield  {title} {\enquote {\bibinfo {title} {Elastic
  properties of self-organized nanogratings produced by femtosecond laser
  exposure of fused silica},}\ }\href {\doibase
  10.1103/PHYSREVMATERIALS.4.023607} {\bibfield  {journal} {\bibinfo  {journal}
  {PHYSICAL REVIEW MATERIALS}\ }\textbf {\bibinfo {volume} {4}},\ \bibinfo
  {pages} {023607} (\bibinfo {year} {2020})}\BibitemShut {NoStop}%
\bibitem [{\citenamefont {Champion}\ and\ \citenamefont
  {Bellouard}(2012)}]{championDirectVolumeVariation2012}%
  \BibitemOpen
  \bibfield  {author} {\bibinfo {author} {\bibfnamefont {A.}~\bibnamefont
  {Champion}}\ and\ \bibinfo {author} {\bibfnamefont {Y.}~\bibnamefont
  {Bellouard}},\ }\bibfield  {title} {\enquote {\bibinfo {title} {Direct volume
  variation measurements in fused silica specimens exposed to femtosecond
  laser},}\ }\href {\doibase 10.1364/OME.2.000789} {\bibfield  {journal}
  {\bibinfo  {journal} {Optical Materials Express}\ }\textbf {\bibinfo {volume}
  {2}},\ \bibinfo {pages} {789--798} (\bibinfo {year} {2012})}\BibitemShut
  {NoStop}%
\bibitem [{\citenamefont {Gleason}\ \emph {et~al.}(2015)\citenamefont
  {Gleason}, \citenamefont {Bolme}, \citenamefont {Lee}, \citenamefont
  {Nagler}, \citenamefont {Galtier}, \citenamefont {Milathianaki},
  \citenamefont {Hawreliak}, \citenamefont {Kraus}, \citenamefont {Eggert},
  \citenamefont {Fratanduono}, \citenamefont {Collins}, \citenamefont
  {Sandberg}, \citenamefont {Yang},\ and\ \citenamefont
  {Mao}}]{gleasonUltrafastVisualizationCrystallization2015}%
  \BibitemOpen
  \bibfield  {author} {\bibinfo {author} {\bibfnamefont {A.~E.}\ \bibnamefont
  {Gleason}}, \bibinfo {author} {\bibfnamefont {C.~A.}\ \bibnamefont {Bolme}},
  \bibinfo {author} {\bibfnamefont {H.~J.}\ \bibnamefont {Lee}}, \bibinfo
  {author} {\bibfnamefont {B.}~\bibnamefont {Nagler}}, \bibinfo {author}
  {\bibfnamefont {E.}~\bibnamefont {Galtier}}, \bibinfo {author} {\bibfnamefont
  {D.}~\bibnamefont {Milathianaki}}, \bibinfo {author} {\bibfnamefont
  {J.}~\bibnamefont {Hawreliak}}, \bibinfo {author} {\bibfnamefont {R.~G.}\
  \bibnamefont {Kraus}}, \bibinfo {author} {\bibfnamefont {J.~H.}\ \bibnamefont
  {Eggert}}, \bibinfo {author} {\bibfnamefont {D.~E.}\ \bibnamefont
  {Fratanduono}}, \bibinfo {author} {\bibfnamefont {G.~W.}\ \bibnamefont
  {Collins}}, \bibinfo {author} {\bibfnamefont {R.}~\bibnamefont {Sandberg}},
  \bibinfo {author} {\bibfnamefont {W.}~\bibnamefont {Yang}}, \ and\ \bibinfo
  {author} {\bibfnamefont {W.~L.}\ \bibnamefont {Mao}},\ }\bibfield  {title}
  {\enquote {\bibinfo {title} {Ultrafast visualization of crystallization and
  grain growth in shock-compressed {{SiO2}}},}\ }\href {\doibase
  10.1038/ncomms9191} {\bibfield  {journal} {\bibinfo  {journal} {Nature
  Communications}\ }\textbf {\bibinfo {volume} {6}},\ \bibinfo {pages} {8191}
  (\bibinfo {year} {2015})}\BibitemShut {NoStop}%
\bibitem [{\citenamefont {Gleason}\ \emph {et~al.}(2017)\citenamefont
  {Gleason}, \citenamefont {Bolme}, \citenamefont {Lee}, \citenamefont
  {Nagler}, \citenamefont {Galtier}, \citenamefont {Kraus}, \citenamefont
  {Sandberg}, \citenamefont {Yang}, \citenamefont {Langenhorst},\ and\
  \citenamefont {Mao}}]{gleasonTimeresolvedDiffractionShockreleased2017}%
  \BibitemOpen
  \bibfield  {author} {\bibinfo {author} {\bibfnamefont {A.~E.}\ \bibnamefont
  {Gleason}}, \bibinfo {author} {\bibfnamefont {C.~A.}\ \bibnamefont {Bolme}},
  \bibinfo {author} {\bibfnamefont {H.~J.}\ \bibnamefont {Lee}}, \bibinfo
  {author} {\bibfnamefont {B.}~\bibnamefont {Nagler}}, \bibinfo {author}
  {\bibfnamefont {E.}~\bibnamefont {Galtier}}, \bibinfo {author} {\bibfnamefont
  {R.~G.}\ \bibnamefont {Kraus}}, \bibinfo {author} {\bibfnamefont
  {R.}~\bibnamefont {Sandberg}}, \bibinfo {author} {\bibfnamefont
  {W.}~\bibnamefont {Yang}}, \bibinfo {author} {\bibfnamefont {F.}~\bibnamefont
  {Langenhorst}}, \ and\ \bibinfo {author} {\bibfnamefont {W.~L.}\ \bibnamefont
  {Mao}},\ }\bibfield  {title} {\enquote {\bibinfo {title} {Time-resolved
  diffraction of shock-released {{SiO}} 2 and diaplectic glass formation},}\
  }\href {\doibase 10.1038/s41467-017-01791-y} {\bibfield  {journal} {\bibinfo
  {journal} {Nature Communications}\ }\textbf {\bibinfo {volume} {8}},\
  \bibinfo {pages} {1481} (\bibinfo {year} {2017})}\BibitemShut {NoStop}%
\bibitem [{\citenamefont {Rajesh}\ and\ \citenamefont
  {Bellouard}(2010)}]{rajeshFastFemtosecondLaser2010}%
  \BibitemOpen
  \bibfield  {author} {\bibinfo {author} {\bibfnamefont {S.}~\bibnamefont
  {Rajesh}}\ and\ \bibinfo {author} {\bibfnamefont {Y.}~\bibnamefont
  {Bellouard}},\ }\bibfield  {title} {\enquote {\bibinfo {title} {Towards fast
  femtosecond laser micromachining of fused silica: {{The}} effect of deposited
  energy.}}\ }\href {\doibase 10.1364/OE.18.021490} {\bibfield  {journal}
  {\bibinfo  {journal} {Optics Express}\ }\textbf {\bibinfo {volume} {18}},\
  \bibinfo {pages} {21490--21497} (\bibinfo {year} {2010})}\BibitemShut
  {NoStop}%
\bibitem [{\citenamefont {Mouskeftaras}\ and\ \citenamefont
  {Bellouard}(2018)}]{mouskeftarasEffectCombinationFemtosecond2018}%
  \BibitemOpen
  \bibfield  {author} {\bibinfo {author} {\bibfnamefont {A.}~\bibnamefont
  {Mouskeftaras}}\ and\ \bibinfo {author} {\bibfnamefont {Y.}~\bibnamefont
  {Bellouard}},\ }\bibfield  {title} {\enquote {\bibinfo {title} {Effect of the
  {{Combination}} of {{Femtosecond Laser Pulses Exposure}} on the {{Etching
  Rate}} of {{Fused Silica}} in {{Hydrofluoric Acid}}},}\ }\href@noop {}
  {\bibfield  {journal} {\bibinfo  {journal} {JLMN-Journal of Laser
  Micro/Nanoengineering}\ }\textbf {\bibinfo {volume} {13}},\ \bibinfo {pages}
  {26--30} (\bibinfo {year} {2018})}\BibitemShut {NoStop}%
\bibitem [{\citenamefont {Agarwal}\ and\ \citenamefont
  {Tomozawa}(1997)}]{agarwalCorrelationSilicaGlass1997}%
  \BibitemOpen
  \bibfield  {author} {\bibinfo {author} {\bibfnamefont {A.}~\bibnamefont
  {Agarwal}}\ and\ \bibinfo {author} {\bibfnamefont {M.}~\bibnamefont
  {Tomozawa}},\ }\bibfield  {title} {\enquote {\bibinfo {title} {Correlation of
  silica glass properties with the infrared spectra},}\ }\href {\doibase
  10.1016/S0022-3093(96)00542-X} {\bibfield  {journal} {\bibinfo  {journal}
  {Journal of Non-Crystalline Solids}\ }\textbf {\bibinfo {volume} {209}},\
  \bibinfo {pages} {166--174} (\bibinfo {year} {1997})}\BibitemShut {NoStop}%
\end{thebibliography}%

\end{document}